\documentstyle[aps,prb]{revtex}
\begin{document}

\title{Ultrasonic attenuation in magnetic fields for 
superconducting states with line nodes in  Sr$_2$RuO$_4$}
\author{L. Tewordt and D. Fay}
\address{I. Institut f\"ur Theoretische Physik,
Universit\"at Hamburg, Jungiusstr. 9, 20355 Hamburg, 
Germany}
\date{\today}
\maketitle
\begin{abstract}
       We calculate the ultrasonic attenuation in magnetic fields 
for superconducting states with line nodes 
vertical or horizontal relative to the RuO$_2$ planes.  This theory,
which is valid for fields near $H_{c2}$ and not too low temperatures,
takes into account the effects of supercurrent flow and Andreev
scattering by the Abrikosov vortex lattice. For rotating in-plane field 
$H(\Theta)$ the attenuation $\alpha(\Theta)$ exhibits  
variations of fourfold symmetry in the rotation angle $\Theta$. In the 
case of vertical nodes, the transverse T100 sound mode yields the 
weakest (linear) $H$ and $T$ dependence of $\alpha$ , while the 
longitudinal L100 mode yields a stronger (quadratic) $H$ and $T$ 
dependence. This is in strong contrast to the case of horizontal line 
nodes where $\alpha$ is the same for the T100 and L100 modes 
(apart from a shift of $\pi/4$ in field direction) and is roughly a 
quadratic function of $H$ and $T$. Thus we conclude that 
measurements of $\alpha$ in in-plane magnetic fields for different 
in-plane sound modes may be an important tool for probing the 
nodal structure of the gap in Sr$_2$RuO$_4$.
\end{abstract}
\vspace{0.5in}
\section{\quad Introduction}

     A number of experiments give evidence that the superconducting 
state in layered Sr$_2$RuO$_4$ \, \cite{Maeno} consists of spin-triplet 
Cooper pairs with broken time-reversal symmetry. More recent
experiments at low temperatures have established power law T
dependence of the specific heat ($T^2$), \, \cite{Nishi} the spin-lattice
relaxation rate ($T^3$), \, \cite{Ishida} the electronic thermal conductivity,
\cite{Izawa,Tanatar} the penetration depth, \cite{Bonalde} and the
ultrasonic attenuation. \cite{Lupien,Matsui} These properties are most
naturally explained in terms of a spin-triplet order parameter 
${\bf d}({\bf p}) = \Delta\,{\bf {\hat{z}}}\,(p_{x}+ip_{y})g({\bf p})$ where
the even-parity function  $g({\bf p})$ has vertical line nodes (e.g.,
$p_{x}^{2}-p_{y}^{2}$ or $p_xp_y$ ) or horizontal line nodes 
[e.g., cos$(cp_z+a_0)\,$] on the cylindrical Fermi surface.  
\cite{Hasegawa,Graf,Dahm} Since the measured anisotropy of
the in-plane thermal conductivity $\kappa[\Theta]$ for rotating 
in-plane magnetic field \, \cite{Izawa} is smaller than the calculated 
anisotropy for vertical nodes, \cite{Dahm} and since the anisotropy
of the inter-plane $\kappa[\Theta]$ is insignificant, \cite{Tanatar} 
these authors have discarded vertical nodes and suggested instead
horizontal nodes in the superconducting gap. Recently we have shown,
however, that the amplitudes of $\kappa[\Theta]$ for vertical and
horizontal line nodes are about the same. \cite{Tew} The small size
of the observed amplitude of $\kappa[\Theta]$ is due to the result that
the amplitude of the variation of $\kappa[\Theta]$ decreases with 
increasing impurity scattering and temperature. The data for ultrasonic
attenuation $\alpha$ in Sr$_2$RuO$_4$ measured for different sound 
directions ${\bf \hat{q}}$ and polarizations ${\bf \hat{e}}$ in the 
ab plane are found to be consistent with a vertical line
node structure $p_{x}^{2}-p_{y}^{2}\,$ or a horizontal line of nodes in
conjunction with significant gap modulation. \cite{Lupien}

      Since the question as to the nodal structure of the superconducting
gap in Sr$_2$RuO$_4$ is still unresolved, we suggest here measurements
of the anisotropy of the ultrasound attenuation $\alpha(\Theta)$ for rotating 
in-plane magnetic fields. Indeed, ultrasonic attenuation is another powerful
tool for probing the anisotropic gap structure because $\alpha(\Theta)$ is
sensitive to the relative orientations of the sound direction and 
polarization, ${\bf \hat{q}}$ and ${\bf \hat{e}}$,  the field ${\bf H}(\Theta)$, 
and the nodal directions of the gap. We shall show that $\alpha(\Theta)$ 
exhibits fourfold symmetric variations in rotation angle $\Theta$ of the 
magnetic field while the variations of $\kappa[\Theta]$ have twofold 
symmetry. It turns out that the field and temperature dependencies of 
$\alpha$ for vertical gap nodes are quite different for longitudinal and 
transverse sound modes while $\alpha$ for horizontal nodes is essentially 
the same for longitudinal and transverse sound modes. Thus observation 
of the field and temperature dependence of $\alpha$ for longitudinal and 
transverse  sound modes should yield important information on the nodal 
structure of the gap in Sr$_2$RuO$_4$. 

      The ultrasonic attenuation $\alpha$ in the vortex state near $H_{c2}$
for type-II s-wave superconductors has been calculated previously by 
Scharnberg \cite{Scharnberg2} and by Klimesh and Pesch. \cite{KlimPesch}
In Ref.~\onlinecite{Scharnberg2} the Green's function $G$
of Brandt, Pesch, and Tewordt (BPT) \cite{BPT} has been employed which 
was derived from the spatial Gorkov integral equation for 
$G({\bf r},{\bf r'},\omega)$ with kernel proportional to the "potential"
\begin{displaymath}
V({\bf r}_1,{\bf r}_2) = \Delta({\bf r}_1)\Delta^{\ast}({\bf r}_2)\, 
exp[-2\imath e \int^{2}_{1}{\bf A}\cdot{\bf {ds}}]
\end{displaymath}
where $\Delta({\bf r})$ is the Abrikosov vortex-lattice order parameter and
${\bf A}$ is the vector potential of the magnetic field. This integral
equation has been solved by expanding all functions in Fourier series 
${\bf k}$ with respect to center of mass coordinates and in Fourier 
integrals ${\bf p}$ with respect to relative (difference) coordinates. In
calculating physical quantities near $H_{c2}$ it often suffices to consider
only the ${\bf k}=0$ Fourier component $G({\bf p},\omega)$. The
corresponding anomalous Green's function $F$ has been derived in the
context of NMR theory by Pesch.\cite{Pesch} The effective self energy
part occuring in the denominators of these Green's functions is 
proportional to the spatial average of $|\Delta({\bf r})|^2$ denoted by
$\Delta^2 \cong  \Delta^2_{BCS}(T)\,(1-H/H_{c2})$. Furthermore it depends
decisively on the quantity $\Lambda/v\sin\theta$ where 
$\Lambda=1/\sqrt{2eH}$ , $v$ is the Fermi velocity in the plane 
perpendiculat to ${\bf H}$, and $\theta$ is the angle between the
quasiparticle momentum ${\bf p}$ and the field ${\bf H}$. For
$\theta\rightarrow0$ the Green's functions $G$ and $F$ tend to the
ordinary Green's functions for an s-wave superconductor with gap $\Delta$.
For finite $\theta$, the self energy contains both the effects of the Doppler
shift and the Andreev scattering by the potential 
$V({\bf p'})=\Delta^2\Lambda^2\delta(p'_z)exp(-\Lambda^2p'^2))$ for
momenta ${\bf p'}$ in the plane perpendicular to ${\bf H}$. This potential
is multiplied by the hole-propagator $G^0({\bf p}-{\bf p'},-\omega)$ and
integrated over all ${\bf p'}$. The theory of $\alpha$ based on the Kubo
formula and the BPT Green's function \cite{Scharnberg2} leads in the 
extreme clean limit to unphysical results in the final expressions. However,
for  small enough mean free paths $\ell$ this method is capable
of describing experiments near $H_{c2}$  sufficiently well.
\cite{HaughtonMaki}

Another approximation scheme for $\alpha$, which is based on the
Eilenberger equations and the Larkin-Ovchinnikov equations for the
correlation functions via linear response theory, \cite{KlimPesch} can be
carried out in the limit $\ell \rightarrow \infty$. In this approximation scheme
only the ${\bf k}=0$ coefficient (spatial average) of the solution is 
considered. These authors state that this approximation is most
questionable for quasiparticle directions ${\bf p}\,||\,{\bf H}$, and therefore
they have concentrated on the special  case of longitudinal ultrasonic
attenuation where the wavevector ${\bf q}$ is parallel to ${\bf H}$. Then
the main contributions to the correlation function arise from directions 
${\bf p}$ perpendicular to ${\bf q}$, and thus perpendicular to ${\bf H}$,
for which their approximation is best.

Since our main aim is to calculate $\alpha$ for unconventional 
superconducting gaps with nodes, we employ BPT Green's
functions  $G$ and $F$ where $\Delta^2$ is replaced by 
$|\Delta\,f({\bf p})|^2$ with $f({\bf p})$ containing the ${\bf p}$-dependence.
This result has been derived from the original spatial integral equation for
$G({\bf r},{\bf r'},\omega)$ which contains in the kernal the non-local order
parameters $\Delta({\bf r}_1,{\bf r'}_1)$ and 
$\Delta^{\ast}({\bf r}_2,{\bf r'}_2)$ .  The method consists in writing
$\Delta({\bf r}_1,{\bf r'}_1) = \Delta({\bf r}_1,{\bf r}_1-{\bf r'}_1)$ and
$\Delta^{\ast}({\bf r}_2,{\bf r'}_2) = \Delta^{\ast}({\bf r}_2,{\bf r}_2-{\bf r'}_2)$ 
and introducing Fourier integrals with respect to the relative coordinates. 
In this way one can show that, to a good approximation
($p' \sim 1/\Lambda \ll p_F$), the $\Delta^2$ in the BPT Green's function
is replaced by $\Delta^2\,|f({\bf p})|^2$. 

Another problem arises from the fact that we consider $\alpha$ for
longitidunal and transverse sound waves with propagation vector ${\bf q}$
in the ab-plane of Sr$_2$RuO$_4$ , together with a field ${\bf H}$ in the
plane which is rotated with respect to the ${\bf {\hat{a}}}$ direction by the 
rotation angle $\Theta$. This means that the angle between ${\bf q}$ and
${\bf H}$ takes on all values, including ${\bf q}$ perpendicular to ${\bf H}$ so
that the main contributions to $\alpha$ arise from directions of ${\bf p}$
near the direction of ${\bf H}$. As has been pointed out above, for these 
directions the BPT Green's function is close to the ordinary Green's
function which is physically quite plausible. Furthermore, for 
${\bf p}\,\,\|{\bf H}$ our correlation function, Eq. (\ref{DefI}), tends to the one 
derived in Ref.~\onlinecite{Scharnberg} for $H=0$. Therefore we believe that, in 
contrast to the semi-classical approximation scheme, \cite{KlimPesch}
our method can deal quite well with general directions of ${\bf H}$ with
respect to ${\bf q}$ in the ab-plane.

In Section II we present the general theory on the basis of the BPT Green's
function. In Section III we discuss the ultrasonic attenuation results for
different sound modes and several superconducting states with vertical
and horizontal line nodes. The conclusions are given in Section IV.

\section{\quad General Theory of Ultrasonic Attenuation near $H_{c2}$}

      Our present theory for the ultrasonic attenuation $\alpha$ closely follows
the method for calculating the thermal conductivity in magnetic fields near
$H_{c2}$. \cite{Tew} We start with the normal and anomalous BPT Green's 
functions $G$ and $F$ which contain both the effects of supercurrent flow
and scattering by the Abrikosov vortex lattice on the quasiparticle 
spectrum. \cite{BPT} These Green's functions are employed in the
expressions for the correlation functions for longitudinal and transverse
sound propagation. Our method for evaluating the Kubo formula follows
closely the method which has been used in the early theory of thermal
conductivity by Ambegaokar and Tewordt. \cite{AmbTew} First the integral of 
$\mbox{Re} [G(\xi,\omega) \cdot G^{\ast}(\xi-{\bf v} \cdot {\bf q},
\omega-\omega_0)-F \cdot F^{\ast}]$ over the energy variable $\xi$ is 
carried out. Here ${\bf q}$ and $\omega_0$ are the wave vector and
frequency of the sound wave and ${\bf v}$ is the Fermi velocity. This yields
the residue of this expression at the pole $\xi = \xi_0$ of the BPT Green's
function $G(\xi,\omega)$ given by Eqs. (\ref{Defs}) and (\ref{transc}). The
most important term in the resulting expression is the quantity 
$\mbox{Im}\,\xi_0$ which yields the scattering rates due to impurity 
scattering and Andreev scattering by the vortices (see Eq. (\ref{DefI})). We 
have obtained the Andreev scattering rate by calculating $\mbox{Im}\,\xi_0$
from Eqs. (\ref{Defs}) and (\ref{transc}) for given $\omega/\Delta$,
 $\Delta\Lambda/v$, and $\gamma=0$ (no impurity scattering) as a 
function of the angle $\tilde{\theta}=\phi-\Theta$ between the quasiparticle
direction ${\bf p}$ and the field ${\bf H}$. Then we find that for 
$\omega \ge \Delta$ (extended states) $\mbox{Im}\xi_0$ is zero in a range
of angles above $\tilde{\theta}=0$ (${\bf p}\,\parallel {\bf H}$) which 
increases with $\omega$, and becomes finite of order $(\Delta\Lambda/v)^2$
in a broad range of angles up to $\tilde{\theta}=\pi/2$ 
(${\bf p}\,\perp {\bf H}$). (See Fig. (6)) It should be pointed out that the
denominator of the second factor in Eq. (\ref{DefI}) (without the absolute
square) occurs also in the expression for the density of states in the
vortex state.\cite{BPT}. In the limit $\tilde{\theta}=\phi-\Theta \rightarrow 0$
(${\bf p}\,\parallel {\bf H}$) the expression for $I$ in Eq. (\ref{DefI}) 
(for $q\ell\rightarrow0$) tends to the well-known expression for the inverse
relaxation rate 
$\mbox{Im}\xi_0\,\rightarrow\,\mbox{Im}(\tilde{\omega}^2-|\Delta|^2)^{1/2}$ 
(where $\tilde{\omega}=\omega+\imath\gamma$) times the coherence
factor $1+(|\tilde{\omega}|^2-|\Delta|^2)/|\tilde{\omega}^2-|\Delta|^2|$ which
was first derived in Appendix C of Ref.~\onlinecite{AmbTew}. In the
hydrodynamic regime, $\omega_0 \tau \ll 1$, we obtain, by including
vertex corrections in analogy to Ref.~\onlinecite{Scharnberg}, the following
expression for the ratio of the ultrasound attenuation in the 
superconducting state, $\alpha_s$, to that in the normal state $\alpha_n$:
\begin{eqnarray}
\frac{\alpha_s}{\alpha_n}
& = &
\int_{0}^{\infty}\frac{d\omega}{2T}\,\mbox{sech}^{2}(\omega/2T)
\left\{  \int_{0}^{2\pi}\frac{d\phi}{2\pi}\: \left[ \pi_{ij}(\phi) \right]^2\:I(\phi)
\right. \nonumber\\
&  & + 
\left. \left[ \int_{0}^{2\pi}\frac{d\phi}{2\pi}\: \left[ \pi_{ij}(\phi) \right] \:I(\phi)
\right]^2 \: \left[ |g|^2 - \int_{0}^{2\pi}\frac{d\phi}{2\pi}\:\  \: I(\phi) \right]^{-1}
\right\} \, ,
\label{alpha}
\end{eqnarray}
where
\begin{equation}
\pi_{xx}^2(\phi)  = 2\,cos^2(2\phi),\quad(\mbox{L100}) \: ;
\quad \pi_{xy}^2(\phi)  = 2\,sin^2(2\phi),\quad(\mbox{T100}) \, ,
\label{Defpi}
\end{equation}
\begin{equation}
I(\phi) = \frac{\mbox{Im}\,\xi_{0}/\Gamma}
{ \left[\,(\mbox{Im}\,\xi_{0}/\Gamma\,)^2+
(q\ell)^2(\hat{q}\cdot\hat{p})^2\right] }\,
\frac{ \left[\,1-\pi[\Delta\Lambda/v\sin(\phi-\Theta)]^{2}\,
|f|^{2}\,|w(z_{0})|^{2}\,\right]}
{|1+2[\Delta\Lambda/v\sin(\phi-\Theta)]^{2}\,|f|^{2}
\left[1+i\sqrt{\pi}z_{0}w(z_{0})\right]|^{2}}\, ,
\label{DefI}
\end{equation}
\begin{eqnarray}
z_{0}=(\omega+i\gamma+\xi_{0}) [\Lambda/v\sin(\phi-\Theta)]\, ,
& &  \qquad\Lambda = (2eH)^{-1/2}\, , \nonumber\\
\gamma = \Gamma / g\, ,\quad & & g = N(\omega,H)/N_0
\label{Defs}
\end{eqnarray}
Here $\pi_{xx}^2$ and $\pi_{xy}^2$ are the weight factors for the
L100 (T110) and T100 (L110) sound modes \, \cite{Lupien,Graf} where
L = longitudinal, T = transverse, and 
${\bf \hat{q}}\parallel[100]\: \mbox{or}\: [110]\,$. We use the following 
notation: $\Gamma$ is the normal-state impurity scattering rate, $v$ 
is the in-plane Fermi velocity, 
$\Theta=\angle({\bf H},{\bf \hat{a}})$ is the direction of ${\bf H}$ in the
ab plane, $\Delta^2$ is the spatial average of the absolute square
of the order parameter for the Abrikosov vortex lattice, and 
$|f(\phi)|^2$ is the normalized absolute square of the gap function.
The quantity $z_0$, and thus the pole $\xi_0$ of $G$, is given by
the transcendental equation: \cite{BPT,Tew}
\begin{equation}
z_{0}=2(\omega+i\gamma)[\Lambda/v\sin(\phi-\Theta)] + 
i\sqrt{\pi}[\Delta\Lambda/v\sin(\phi-\Theta)]^{2}\,|f|^{2}\,w(z_{0})\, .
\label{transc}
\end{equation}
Here, $w(z)=exp(-z^2)erfc(-\imath z)$. Note that Eqs. (\ref{Defs}) and
(\ref{transc}) differ from the corresponding equations in Ref.~\onlinecite{BPT}
in that the $\sin\theta$ in the original quantity $\Delta\Lambda/v \sin\theta$
is replaced by $\sin(\phi-\Theta)$ for ${\bf p}$ lying in the direction $\phi$
and ${\bf H}$ lying in the direction $\Theta$ in the ab-plane, and $\Delta^2$
 is replaced by $\Delta^2 |f(\phi)|^2$ for a gap with vertical lines of nodes. 
For a general state with $f({\bf p})$ one has to replace $\sin\theta$ with
the expression given in Eq. (7) of Ref.~\onlinecite{Tew} and carry out the 
double integral over the polar and azimuthal angles $\theta$ and
$\phi$. In the following calculations we will neglect the term proportional
to $(q \ell)^2$ in the first denominator of Eq. (\ref{DefI}) assuming the long
wavelength limit $q \ell \ll1$. The dependencies of $(\Delta\Lambda/v)^2$ 
and $(\Lambda/v)^2$ on
$H/H_{c2}$ are presented in Ref.~\onlinecite{Tew}. The expression for
ultrasonic attenuation in Eq. (\ref{alpha}) is similar to the expression for
$\kappa$ in  Ref.~\onlinecite{Tew} apart from the missing factor 
$\omega^2$, the weight factor $\pi_{ij}^2$ instead of $v_i^2$, and the
vertex corrections [second term in the curly brackets in Eq. (\ref{alpha})]. 
For the vertex corrections we have used the expressions 
in Ref.~\onlinecite{Scharnberg} for the unitary limit $\delta_N = \pi / 2$ of
 the phase shift for impurity scattering. In the limit 
$\omega\rightarrow 0\,$, an analytical expression for the solution
$z_0 = ix_0$ of Eq. (\ref{transc}) is obtained which yields, in close
analogy to $\kappa\,$, \cite{Tew} a much simpler expression for 
$\alpha_s / \alpha_n$ in the limit $T \rightarrow 0$.

\section{\quad Results and discussion of ultrasonic attenuation for
different sound modes and several superconducting states with vertical
and horizontal line nodes}

     First we consider the $f$-wave pairing state 
${\bf d} = \Delta\,{\bf {\hat{z}}}\,(p_{x}+ip_{y})(p_{x}^{2}-p_{y}^{2})$
for which $|f|^2 = \cos^2(2\phi)$ has four vertical line nodes at
$\phi = \pi/4,\,3\pi/4,\, \cdots$ on the cylindrical Fermi surface. In Fig. 1
we show our results for $\alpha_s / \alpha_n$ versus $H/H_{c2}$ 
for the L100 ($\Theta=0$) and T100 ($\Theta=\pi/4$) modes at $T=0$
and impurity scattering rate $\Gamma/\Delta_0 = 0.1$. $\Delta_0$ is
the BCS gap parameter, and we will always use the value 
$\beta_A = 1.2$ for the Abrikosov parameter. It is seen that 
$\alpha_s / \alpha_n$ for the L100 mode exhibits a strong upward
curvature near $H_{c2}$ while it is a more linear function of $H$ for the
T100 mode. For increasing $\Gamma/\Delta_0$ the upward curvatures
decrease. 

     The solid curves for $\alpha_s / \alpha_n$ in Fig. 1 are calculated
by neglecting the vertex corrections in Eq. (\ref{alpha}), while the
dashed curves include the vertex corrections in the unitary impurity
scattering limit. One recognizes that the effect of the vertex corrections
in the unitary limit is rather small. This is also true for impurity
scattering in the Born limit where $|g|^2$ in the denominator of 
Eq. (\ref{alpha}) is replaced by one. It is interesting
that, for field direction angles $\Theta=\pi/4$ and $\Theta=0$, the
vertex corrections for the L100 and T100 modes, respectively, vanish.
In the following calculations and figures we shall neglect the vertex 
corrections.

      In Fig. 2 we show the dependence of $\alpha_s / \alpha_n$
on the in-plane field direction $\Theta$ at $T=0$ for fixed field strength,
$\Delta\Lambda/v = 0.2$, and impurity scattering rate 
$\Gamma/\Delta_0 = 0.1$. One sees that $\alpha(\Theta)$ has 
variations in $\Theta$ of fourfold symmetry where the minima and
maxima occur at $\Theta = 0$ and $\pi / 4$ for the L100 mode. For the
T100 mode the maxima and minima are reversed. The maxima and
minima can be explained as an effect of the $\phi$ dependence
of the weighting factors $\pi_{ij}^2$ in Eq. (\ref{Defpi}) and the density 
of states since the expression $|\cdots|^{-2}$ in the denominator of 
$I(\phi)$ in Eq. (\ref{DefI}) is
proportional to $|N(\phi,\Theta)|^2\,$. The density of states has
a minimum (maximum) for quasiparticles traveling parallel (perpendicular)
to the field. \cite{BPT} The minima for L100 and T100 occur at  
$\Theta = 0$ and $\pi / 4$, respectively, because the corresponding 
weight factors  $\pi_{xx}^2$ and $\pi_{xy}^2$ have maxima at 
$\Theta = 0$ and $\pi / 4$. For higher fields (e.g., 
$\Delta\Lambda/v = 0.1$) a small local maximum and two neighboring 
minima occur around $\Theta=\pi/4$ for the T100 mode due to the node of the
gap at $\Theta=\pi/4$.

      The sound atteunation for the other $f$-wave pairing state, 
proportional to $p_xp_y$ with $|f|^2=\sin^2(2\phi)\,$, is obtained from the
previous expression for $|f|^2=\cos^2(2\phi)$ by a simple transformation,
$\phi' = \phi - \pi/4$, in Eq. (\ref{alpha}). This exchanges the results for L100 
and T100 which become functions of the new field rotation angle 
$\Theta' = \Theta - \pi/4$.

       We consider now the spin-triplet pairing state with horizontal line
nodes \, \cite{Hasegawa} where the squared gap amplitude is
proportional to $|f|^2=\cos^2(c\,p_z) = \cos^2[\chi]\,$. Then, in addition
to the $\phi$ integration in Eq. (\ref{alpha}), one also has to do the
integration over $\chi$ from $-\pi$ to $+\pi$. In Fig. 1 we have included
our result for $\alpha_s / \alpha_n$ versus $H/H_{c2}$ for the 
mode L100 at $T=0$ and $\Gamma/\Delta_0 = 0.1$. A strong upward
curvature occurs near $H_{c2}$ which is similar to that of the L100 mode
for vertical nodes, but is quite distinct from the almost linear dependence
of $\alpha_s / \alpha_n$ on $H/H_{c2}$ for the T100 mode in the case
of vertical nodes. In Fig. 2 we have plotted our results for the
in-plane field variation $\alpha(\Theta)$ for the L100 and T100 modes.
Note that the function $\alpha(\Theta)$ for T100 is obtained by shifting 
the function for L100 by $\Theta = \pi / 4$ along the $\Theta$ axis. This
can be seen by a variable transformation $\phi' = \phi - \pi/4$
in Eqs. (\ref{alpha})--(\ref{DefI}) making use of the fact that 
$|f|^2=\cos^2(\chi)$ does not
depend on $\phi$. Comparison of the curves in Fig. 2 shows that the 
amplitudes of the variations $\alpha(\Theta)$ for vertical and
horizontal nodes are about the same (here about 10\%). However, there
is a marked difference in the form of the variations for the L100 and
T100 modes for vertical nodes because, for the T100 mode, the gap
node manifests itself in the structure around $\pi / 4$. In contrast to
this, the form of the variations for horizontal nodes is the same apart
from the shift by $\pi/4$ along the $\Theta$ axis.

     We turn now to the $\omega$ dependence of the $\phi$ integrals in 
Eq. (\ref{alpha}) which determine the T dependence of 
$\alpha_s / \alpha_n\,$. It is now necessary to solve Eq. (\ref{transc}) 
for $z_0$ as a function of $\Omega = \omega/\Delta\,$, and then to
integrate the resulting expression for $I(\phi)$ in Eq. (\ref{DefI}) 
(multiplied by $\pi_{ij}^2(\phi)$, etc.)
over $\phi$ (and $\chi$ for horizontal nodes). We denote 
by $\alpha_s(\Omega)/\alpha_n\,$ the resulting
integrand of the $\omega$ integral [without the factor 
$(1/2T)\mbox{sech}^{2}(\omega/2T)$]. In Fig. 3(a) we plot 
$\alpha_s(\Omega)/\alpha_n\,$ vs $\Omega$ for 
$|f|^2= \cos^2(2\phi)$ for large field magnitude 
[$\Delta\Lambda/v = 0.2\:(H/H_{c2}=0.78)$]
and field directions $\Theta=0$ and $\pi/4$ for both the L100 and T100 modes.
It is interesting that the curves for $\Theta = 0$ and $\pi/4$ cross at finite
frequencies $\Omega$ which means that the minima and maxima in
$\alpha(\Theta)$ are interchanged. Furthermore, the $\Omega$
dependence for T100 at $\Theta=\pi/4$ is almost linear while, for L100 and
$\Theta=0$, it is quadratic. In Fig. 3(b) we show
$\alpha_s(\Omega)/\alpha_n$ for the state $|f|^2= \cos^2(cp_z)\,$.

     The temperature dependence of $\alpha_s/\alpha_n$ is obtained
according to Eq. (\ref{alpha}) by integrating the expression
$(1/2T)\mbox{sech}^{2}(\omega/2T)\,\alpha_s(\Omega)/\alpha_n$ over
$\omega\,$. For the variable transformation from $\omega$ to 
$\Omega=\omega/\Delta$ we make use of the relations
$\Delta/\Delta_0=[1-(H/H_{c2})]^{1/2}/\sqrt{\beta_A}$ and
$H/H_{c2} = [1+6\beta_A(\Delta\Lambda/v)^2]^{-1}\:$. \cite{Tew} In Fig. 4
we show our results for $\alpha_s/\alpha_n$ vs $T/T_c$ for vertical gap
nodes at low and high field strength ($\Delta\Lambda/v = 0.6$ and $0.2$,
or $H/H_{c2}= 0.28$ and $0.78$) and field directions $\Theta=0$ and 
$\Theta=\pi/4$ for the L100 and T100 modes. One recognizes that, for low
$T$, the attenuation of the L100 mode is approximately a quadratic
function of $T$, while it is nearly linear for the T100 mode at 
$\Theta=\pi/4$. These different low $T$ power laws for the L100 and 
T100 modes agree roughly with the results of Ref.~\onlinecite{Graf} for 
zero field. It is interesting that the quadratic and linear $T$ dependencies 
of $\alpha_s/\alpha_n$ for the L100 and T100 modes correspond to the 
quadratic and linear dependencies of $\alpha_s/\alpha_n$ on
$H/H_{c2}$ for these modes (see Fig. 1). In Fig. 4 we also show
$\alpha_s/\alpha_n$ vs $T/T_c$ for the L100 mode for horizontal gap 
nodes and field directions $\Theta = 0$ and $\pi/4\,$. We recall that the
results for L100 for $\Theta=0\:(\pi/4)$ are identical to those for T100 for
$\Theta=\pi/4\:(0)\,$. One sees that the dependence on $T$ is nearly 
quadratic.

\section{\quad Conclusions}

     In summary, we have calculated the ultrasonic attenuation $\alpha$
in layered Sr$_2$RuO$_4$ in the presence of in-plane magnetic fields 
for spin-triplet superconducting states with vertical or horizontal line
nodes in the gap. This theory takes into account the effects of
supercurrent flow and Andreev scattering (see Fig. (6)) by the Abrikosov
vortex lattice near $H_{c2}$. For rotating in-plane field $H(\Theta)$ the 
attenuation $\alpha(\Theta)$ exhibits variations of fourfold symmetry in the
rotation angle $\Theta$. At $T=0$ the minima occur at $\Theta=0$ and
$\pi/2$ for the L100 sound mode, and at $\Theta=\pi/4$ and $3\pi/4$ for
the T100 mode. The amplitudes of the variations are about about the 
same ($\approx10\%$) for both horizontal and vertical nodes. However, in
the case of vertical nodes, $\alpha(\Theta)$ for the T100 mode shows
characteristic structure in the directions of the nodes. For horizontal nodes 
the variations $\alpha(\Theta)$ for the L100 and T100 modes are the same
if $\Theta$ is shifted by $\pi/4$. The distinction between vertical and
horizontal nodes also manifests itself in the different field and temperature
dependencies of $\alpha$. For vertical nodes the ratio $\alpha_s/\alpha_n$
exhibits a strong upward curvature as a function of $H/H_{c2}$ near 
$H_{c2}$ for the L100 mode while it is more linear for the T100 mode. In the
case of horizontal nodes, the attenuation for both the L100 and T100 modes 
shows a strong upward curvature near $H_{c2}$. The temperature
dependence of the sound attenuation is determined by the frequency
dependence of the integrand $\alpha(\Omega)$, $\Omega=\omega/\Delta$,
in the expression for $\alpha_s/\alpha_n$. It is interesting that the curves for
the functions $\alpha(\Omega)$ for field directions $\Theta=0$ and $\pi/4$
cross each other twice for increasing $\Omega$ indicating that the maxima
and minima are interchanged twice. This causes corresponding crossings
in the $T$  dependencies of $\alpha$ for the two field directions. For vertical 
gap nodes, at low $T$, $\alpha_s/\alpha_n$ exhibits approximately a $T^2$
power law for the L100 mode and a roughly linear T dependence for the T100
mode.  For horizontal gap nodes the functions $\alpha_s/\alpha_n$ vs 
$T/T_c$ for the L100 and T100 modes are identical for field direction differing
by $\pi/4$. For low and high fields $\alpha_s/\alpha_n$ follows
roughly a $T^2$ power law.

      The ultrasonic attenuation in Sr$_2$RuO$_4$ in zero field has been 
measured for the four in-plane modes L100, T110, T100, and L110. 
\cite{Lupien} The attenuation  follows a low temperature power law 
$T^{1.8}$ for the L100 (T110) mode and $T^{1.4}$ for the T100 mode.
In Ref.~\onlinecite{Matsui} a $T^2$ power law below $T_{imp}$ and a
$T^3$ dependence above $T_{imp}$ has been measured for the T110
mode. Calculations based on the assumption of a circular cylindrical
Fermi surface and vertical gap nodes \, \cite{Graf} yield a linear $T$
dependence of $\alpha_s/\alpha_n$ for the T100 (L110) mode which
disagrees with the measured $T^{1.4}$ power law and, for the L100 mode,
a power law close to the measured one. The fact that the weakest T
dependence occurs for the T100 mode is taken as an indication \,
\cite{Lupien} that the gap nodes point in the [110] direction because this
yields an excess of quasiparticles with wavevectors in this direction.

    Our results for the ultrasound attenuation for vertical nodes in 
magnetic fields up to $H_{c2}$ also yield the weakest (linear) $H$
and $T$ dependence for the T100 mode because the corresponding
weight factor has a maximum in the direction of a vertical gap node.
Our calculations yield a stronger (quadratic) $H$ and $T$ 
dependence for the L100 mode because the corresponding weight 
factor has a maximum in the direction of a vertical gap antinode.
This is in strong contrast to the case of horizontal line nodes where
we obtain a roughly quadratic $H$ and $T$ dependence for {\it{both}}
the L100 and T100 modes, which are actually identical for in-plane
field directions differing by $\pi/4$. 

     At this point we would like to mention the modifications required
if one takes the model of horizontal line nodes suggested in 
Ref.~\onlinecite{Lupien}. This model can be simulated by multiplying the 
cos$(cp_z)$ used in this paper (see Ref.~\onlinecite{Hasegawa}) by a gap
with fourfold angular modulation, 
$(\,1+\lambda\mbox{cos}4\phi\,)$, $0<|\lambda|<1$, in the plane. 
This has the effect that the dashed curves in Fig. 2 for the variations
$\alpha(\Theta)$ of the L100 and T100 modes drift apart with a shift 
(and thus the anisotropy) that increases with increasing 
amplitude $\lambda$ of the fourfold gap modulation. See Fig. 5a for the
case $\lambda=1/2$. Corresponding 
splittings occur in the attenuation curves $\alpha$ versus the 
field (Fig. 1) and temperature (Fig. 4)  for the L100 (T110) and T100 
modes which are identical in our simple model of horizontal line 
nodes (see Fig. 5b). In fact, the $H$ and $T$ dependencies of $\alpha$ 
for the T100 mode become weaker than those of the L100 mode for
increasing amplitude $\lambda$ of the gap modulation. We remark that
such a fourfold angular modulation occurs in the gaps of the passive
$\alpha-$ and $\beta-$bands where horizontal line nodes are induced
by the nodeless p-wave gap in the active band by interband proximity
effect.\cite{ZhitRice} We have fitted this gap modulation in the passive
bands with our gap model discussed above and estimate a rather small
value for $\lambda$ of about 0.1. The p-wave gap without line nodes
is approximated in this model with $\lambda=-0.69$. \cite{MiyNar} For
this p-wave gap the attenuation curves for the T100 and L100 modes
in Figs. 1 and 2 (solid curves) are, roughly speaking, interchanged,
which is plausible because the gap minima occur in the directions 
$\phi = 0,\,\pi/2, \ldots$ . This has the effect that the $H$ and $T$
dependencies of $\alpha$ for the L100 mode become weaker than
those of the T100 mode. This appears to be in disagreement with 
experiment. \cite{Lupien} 

      It should be pointed out that our theory is based on a number
of simplifying assumptions.  First, we have considered only the 
lowest order Fourier coefficient of the Green's functions with 
respect to the reciprocal Abrikosov vortex lattice.  As the
field decreases, the higher Fourier coefficients become more and
more important and, near $H_{c1}$, the Doppler shift method for
a single vortex becomes more appropriate. Second, we have assumed
a cylindrical single sheet Fermi surface, whereas the huge 
anisotropy observed between the T110 and T100 normal state
attenuation coefficients \, \cite{Lupien} indicates that the 
anisotropy follows from the nature of the three orbitals forming
the three sheets of the Fermi surface in Sr$_2$RuO$_4$.
\cite{Walker} These authors showed that in the case of 
Sr$_2$RuO$_4$ a simple viscosity tensor describing the
electron-phonon interaction may not be sufficient for ultrasound
attenuation. If this is the case our predictions could be changed. Third,
we have assumed that the attenuation coefficients normalized to their
normal state values are insensitive to the observed normal state 
anisotropies. Here we have taken a temperature independent
scattering rate $\Gamma$ giving rise to a temperature independent
normal state attenuation which agrees roughly with the data for the
L100 and T100 modes. \cite{Lupien}  Thus it cannot be excluded 
that the effect of the normal state anisotropy exceeds all anisotropies
in the superconducting state calculated in this work.

     In spite of the above uncertainties we believe that measurements
of the field dependencies (in addition to the temperature dependencies)
of the ultrasonic attenuation for different in-plane sound modes can
still provide useful information about the nodal structure of the energy
gap in Sr$_2$RuO$_4$.

\acknowledgements
 
We thank K. Scharnberg, N. Schopohl, and T. Dahm for helpful discussions. 
\newpage
\newpage
\vspace{0.5in}
FIGURE\ CAPTIONS \\
1. Ultrasonic attenuation $\alpha_s/\alpha_n$ vs applied field $H/H_{c2}$
at $T=0$ for longitudinal and transverse in-plane sound modes 
L100 ($\Theta = 0$; $\Theta=\angle({\bf H},{\bf \hat{a}})$) and T100
($\Theta = \pi/4$) for impurity scattering rate $\Gamma/\Delta_0 = 0.1$.
Vertical gap nodes: solid and dashed curves without and with vertex
corrections in the unitary impurity scattering limit, respectively.
Horizontal gap nodes: dash-dot curve.

2. $\alpha_s/\alpha_n$ vs $\Theta=\angle({\bf H},{\bf \hat{a}})$ for in-plane
field rotation at $T=0$ for sound modes L100 and T100 and gap parameter
$\Delta\Lambda/v = 0.2\:(H/H_{c2}=0.78)$ and $\Gamma/\Delta_0 = 0.1$.
Solid curves: vertical gap nodes. Dashes curves: horizontal gap nodes.

3. Integrand $\alpha_s(\Omega)/\alpha_n$ in Eq. (\ref{alpha}) for 
$\alpha_s/\alpha_n$ (without the factor $(1/2T)\mbox{sech}^{2}(\omega/2T)$)
vs reduced frequency $\Omega=\omega/\Delta$ for gap parameter 
$\Delta\Lambda/v = 0.2\:(H/H_{c2}=0.78)$ and field directions $\Theta=0$
(solid curves) and $\Theta=\pi/4$ (dashed curves), and 
$\Gamma/\Delta_0 = 0.1$.  (a) L100 and T100 for vertical gap nodes; 
(b) L100 for horizontal gap nodes.

4. $\alpha_s/\alpha_n$ vs $T/T_c$ for vL100, vT100, and hL100 (v = vertical, 
h = horizontal gap nodes), for field directions $\Theta=0$ (solid curves) and 
$\Theta=\pi/4$ (dashed curves). 
(a) $\Delta\Lambda/v = 0.6\quad(H/H_{c2}=0.28)$, $\Gamma/\Delta_0 = 0.2$;
(b) $\Delta\Lambda/v = 0.2\quad(H/H_{c2}=0.78)$, $\Gamma/\Delta_0 = 0.1$.

5. $\alpha_s/\alpha_n$ for a gap with horizontal line nodes and fourfold
angular modulation in the plane ($\lambda=1/2$) for the L100 (solid curves) 
and T100 (dashed curves) sound modes. 
(a) $\alpha_s/\alpha_n$ vs $\Theta=\angle({\bf H},{\bf \hat{a}})$  for
$H/H_{c2}=0.78$;  (b) $\alpha_s/\alpha_n$ vs $H/H_{c2}$ at $T=0$ and for
$\Gamma/\Delta_0 = 0.1$.  

6. Andreev scattering given by $\mbox{Im}\,\xi_0/\Delta_0$ 
(for $\gamma=0$),
versus $\tilde{\theta} = \angle({\bf p},{\bf H})$ for fixed $\tilde{\Delta}
=\Delta\Lambda/v\approx[(H_{c2}/H)-1]^{1/2}/\sqrt{6\beta_A}$ and 
$\Omega=\omega/\Delta$. (a) $\tilde{\Delta}=0.2$ and $\Omega=$0.0, 1.0,
1.2, and 1.5 (from top to bottom). (b) $\Omega=1.0$ and $\tilde{\Delta}=$
0.6, 0.2, and 0.1 (from top to bottom).  (c) $\Omega=1.2$ and 
$\tilde{\Delta}=$
0.6, 0.2, and 0.1 (from top to bottom).


\begin{references}
%
%
\bibitem{Maeno}Y. Maeno et al., Nature (London) {\bf 372},  
532 (1994).
%
\bibitem{Nishi}S. NishiZaki, Y. Maeno, and Z. Mao,  J. Low Temp. Phys. 
{\bf 117}, 1581 (1999); J. Phys. Soc. Jpn. {\bf 69}, 572 (2000).
%
\bibitem{Ishida}K. Ishida et al., Phys. Rev. Lett. {\bf 84}, 5387 (2000).
%
\bibitem{Izawa}K. Izawa et al., Phys. Rev. Lett. {\bf 86}, 2653 (2001).
%
\bibitem{Tanatar}M. A. Tanatar et al., Phys. Rev. Lett. {\bf 86}, 2649 (2001).
%
\bibitem{Bonalde}I. Bonalde et al., Phys. Rev. Lett. {\bf 85}, 4775 (2000).
%
\bibitem{Lupien}C. Lupien et al., Phys. Rev. Lett. {\bf 86}, 5986 (2001).
%
\bibitem{Matsui}H. Matsui et al., Phys. B {\bf 63}, 060505(R).
%
\bibitem{Hasegawa}Y. Hasegawa, K. Machida, and M. Ozaki, 
J. Phys. Soc. Jpn. {\bf 69}, 336 (2000).
%
\bibitem{Graf}M. J. Graf and A. V. Balatsky, Phys. B {\bf 62}, 9697 (2001).
%
\bibitem{Dahm}T. Dahm, H. Won, and K. Maki, cond-mat/0006301 (unpublished).
%
\bibitem{Tew}L. Tewordt and D. Fay, Phys. Rev. B {\bf 64}, 24528 (2001).
%
\bibitem{Scharnberg2}K. Scharnberg, J. Low Temp. Phys. {\bf 6}, 51 (1972).
%
\bibitem{KlimPesch}P. Klimesch and W. Pesch, J. Low Temp. Phys. {\bf 32}, 
869 (1978).
%
\bibitem{BPT}U. Brandt, W. Pesch, and L. Tewordt, Z. Phys. {\bf 201}, 
209  (1967).
%
\bibitem{Pesch}W. Pesch, Phys. Lett. {\bf 28}, 71 (1968); Ph.D. thesis, 
Hamburg,1968 unpublished.
%
\bibitem{HaughtonMaki}A. Houghton and K. Maki, Phys. Rev. B {\bf 4},
843 (1971).
%
\bibitem{Scharnberg}K. Scharnberg, D. Walker, H. Monien, L. Tewordt,
and R. A. Klemm, Solid State Commun. {\bf 60}, 535 (1986).
%
\bibitem{AmbTew}V. Ambegaokar and L. Tewordt, Phys. Rev. {\bf 134},
A805 (1964).
%
\bibitem{ZhitRice}M. E. Zhitomirsky and T. M. Rice, Phys. Rev. Lett. {\bf 87},
057001 (2001).
%
\bibitem{MiyNar}K. Miyake and O. Narikiyo, Phys. Rev. Lett. {\bf 83}, 
1423 (1999).
%
\bibitem{Walker}M. B. Walker, M. F. Smith, and K. V. Samokhin, 
cond-mat/0105109 (unpublished).
%
\end{references}
\end{document}